\documentclass{JFM-FLM_Clean}
\usepackage{anyfontsize}

\usepackage{newtxtext}
\usepackage{newtxmath}

\usepackage{siunitx}
\DeclareSIUnit\mmHg{mmHg}

\usepackage{bm} 
\usepackage{mathrsfs} 
\allowdisplaybreaks
\usepackage{cancel}
\usepackage{accents}

\usepackage{multirow}
\usepackage{booktabs}

\newcommand*{\defeq}{\stackrel{\text{def}}{=}}

\usepackage{natbib}
\usepackage[doipre={doi:}]{uri}
\DeclareUrlCommand\doi{\def\UrlLeft##1\UrlRight{\href{http://dx.doi.org/##1}{doi:##1}}\urlstyle{rm}}

\usepackage{orcidlink}

\usepackage{hyperref}
\hypersetup{
    colorlinks = true,
    urlcolor   = blue,
    linkcolor  = blue,
    citecolor  = blue,
    hypertexnames = false,
}

\lefttitle{N.\ Surianarayanan and I.\ C.\ Christov}
\righttitle{Peristalsis in compliant porous eccentric annuli}

\title{Flow in a porous non-axisymmetric annular conduit: Coupling wall compliance and peristalsis}
\author{Nishanth Surianarayanan\aff{1}\orcidlink{0009-0002-9717-0155} and Ivan C.\ Christov\aff{1}\orcidlink{0000-0001-8531-0531}}

\corresau{Ivan C.\ Christov, \email{christov@purdue.edu}}

\affiliation{\aff{1}School of Mechanical Engineering, Purdue University, West Lafayette, Indiana 47907, USA}

\begin{document}

\maketitle

\begin{abstract}
Coenen \textit{et al.}\ (\textit{J. Fluid Mech.}, vol.~921, 2021, p.~R2) developed a reduced-order model of peristaltic pumping in non-axisymmetric annular conduits with rigid walls, in the context of periarterial space (PAS) flows. \textit{In vivo} studies show that the PAS's outer wall undergoes significant displacement due to flow within and that the penetrating PASs form a porous pathway. To account for these biomechanical aspects, we revisit the problem of flow in an eccentric annular conduit and incorporate porous drag and two-way-coupled fluid--structure interaction between the compliant outer wall and the cerebrospinal fluid flow within. A Darcy--Brinkman term in the axial momentum equation accounts for drag due to the porous medium. We account for changes in hydraulic resistance due to peristalsis and compliant-wall displacements perturbatively, thereby reducing the problem to a single nonlinear partial differential equation for the axial pressure. This reduced-order model allows us to build a mechanistic understanding of flow through a porous penetrating PAS and enables parametric studies. For small-amplitude peristaltic waves, analytical solutions are possible.
\end{abstract}


\section{Introduction}
\label{sec:intro}

Peristalsis is a common driving mechanism of fluid flow within our bodies, caused by pulsations that propagate along the walls of conduits \citep{Jaffrin1971}. This transport mechanism has been observed in the gastrointestinal tract \citep{Acharya2022}, in the urinary system \citep{Cummings2004}, in the reproductive tract \citep{Leftwich2024}, and in lymphatic vessels \citep{Winn2024}. The glymphatic system \citep{Iliff2012A} provides an intriguing case in which peristaltic transport may arise in complex geometry.

The glymphatic system involves the flow of cerebrospinal fluid (CSF) driven by arterial pulsations within annular conduits called the periarterial space (PAS) \citep{Iliff2012A,Mestre2018,Hauglund2025}. Several features have been observed in \textit{in vivo} experimental studies. The first feature is that the arteries are often placed off-center within the PASs \citep{Tithof2019}. The second feature inferred from \textit{in vivo} flow tracer studies is that the PASs located on the brain surface (pial PAS) are open spaces and those penetrating within the brain tissue (penetrating PAS) are porous \citep{Bedussi2018, Mestre2018,Iliff2012A, Schain2017, Bedussi2017}. This observation was also supported by the recent PAS morphology characterization study by \cite{Mestre2022}, which showed that the merging of the epipial and/or intimal pial layers at the entrance of penetrating arterioles formed a porous structure that restricted the entry of large tracer particles. The third feature noted is that the astrocyte endfeet layer is displaced by flow within the PAS, which, in turn, can affect hydraulic resistance in these narrow spaces, thereby presenting a two-way-coupled biomechanical fluid--structure interaction \citep{HH11}. Endfeet-layer displacements of $\approx 1$--$1.5~\si{\micro\meter}$ and $\approx 2~\si{\micro\meter}$ have been observed during functional hyperemia \citep{Holstein-Rnsbo2023,Kedarasetti2020} and natural sleep--wake variation \citep{Bojarskaite2023}, respectively. Given these features, understanding the mechanics of PAS flows requires integrating the principles of peristalsis, transport in porous media, and fluid--structure interactions.

To address the first feature of flows in the PAS, \cite{Tithof2019} investigated the hydraulic resistance to steady flow through a non-porous annular PAS, focusing on the area ratios, aspect ratios, and inner cylinder eccentricities. They determined the resistance by numerically solving the axial momentum equation, subject to no-slip boundary conditions at the inner and outer walls of the PAS. Meanwhile, \cite{Carr2021} performed numerical studies of the flow driven by sinusoidal waves in open eccentric PAS geometries of domain length equal to twice the wavelength. In parallel, \cite{Coenen2021} developed a reduced-order model based on the lubrication approximation to describe flows driven by waves of arbitrary shape in open eccentric PAS. However, both of these works assumed a rigid outer wall. 
To address the second and third features of flows in the PAS, \cite{Kedarasetti2020} performed computational fluid--structure interaction simulations of peristalsis in a porous, penetrating PAS domain and showed that modeling the outer wall as rigid led to unrealistic pressure values, highlighting the importance of modeling the outer wall as compliant. Meanwhile, \cite{Romano2020} and \citet{Gan2023} developed lubrication models for peristaltic pumping in a perivascular space domain connected to a porous brain tissue domain through a permeable outer wall (i.e., astrocyte endfeet layer). However, both of these works idealize the PAS conduits as concentric. \citet{Takagi2024} considered peristaltic pumping in a non-axisymmetric compliant conduit, specifically an eccentric annulus with porous \emph{walls}, but focused on dynamics relevant to marine organisms.

To our knowledge, no two-way-coupled fluid--structure interaction model is currently available for peristalsis in narrow, porous, non-axisymmetric annular conduits under conditions relevant to CSF flows in the PAS. In this work, we address this knowledge gap. In \S\ref{sec:methods}, we derive the governing equations under the lubrication approximation. We employ a perturbation approach to account for changes in resistance due to the propagation of traveling waves along the inner wall and for displacements at the outer wall caused by elastic deformation. The result is a reduced-order two-way-coupled model in the form of a single nonlinear unsteady partial differential equation (PDE) for the axial pressure variation. In \S\ref{sec:Results}, we demonstrate our model on the example case of peristaltic pumping driven by cardiac pulsations in a porous non-axisymmetric penetrating PAS, focusing on understanding how the hydrodynamic resistance, pressure, and flow are affected by eccentricity, the Darcy number, and the compliance number. We conclude in \S\ref{sec:Conclusions}.

\section{Mathematical model under the lubrication approximation}
\label{sec:methods}

\begin{figure}
    \centering
    \includegraphics[width=1\textwidth]{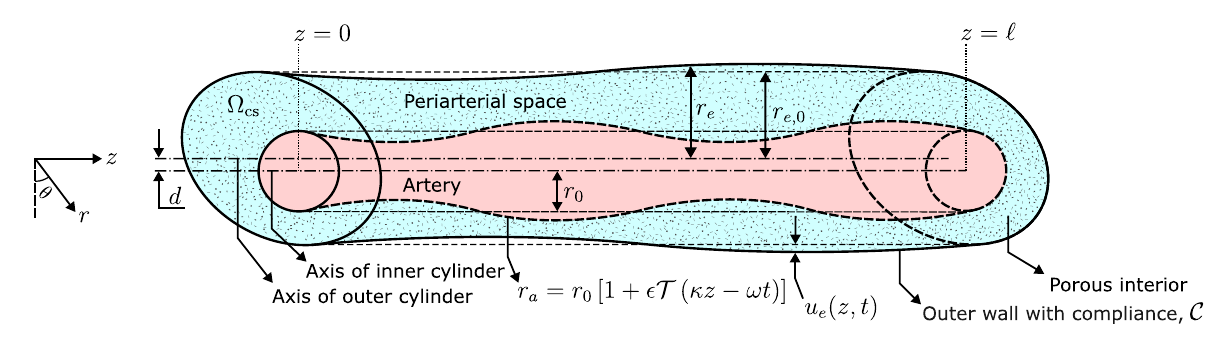}
    \caption{Schematic of the model porous domain, with non-axisymmetric cross-section $\Omega_{\mathrm{cs}}$, and associated mathematical notation. The boundary of each cross-section, $\partial \Omega_{\mathrm{cs}}$, consists of the outer wall (``endfeet layer'' with radius $r_e$ with respect to its own axis), which is always centered at the axis, and the inner cylinder (``artery'' with radius $r_a$ with respect to its own axis), which is offset from the former's axis, leading to a non-axisymmetric space. The offset between the axes of the two walls is $d$ along the ray $\theta = 0$.}
    \label{fig:PVS_ug_schematic}
\end{figure}

Consider a flow in a finite-length annular conduit with a porous interior and a compliant outer wall as shown in figure~\ref{fig:PVS_ug_schematic}. The instantaneous inner and outer radii of the annular conduit are denoted by $r_a$ and $r_e$, with corresponding equilibrium values (i.e., in the absence of peristalsis or deformation) being $r_0$ and $r_{e,0}$, attained at the reference pressure $p_\mathrm{ref}$. Note that expressing the inner and outer boundaries of the PAS in the given cylindrical coordinate system requires care. Peristalsis causes the inner radius, measured about the artery's own axis, to vary as
\begin{equation}
    r_a(z,t) = r_0 [1+\epsilon \mathcal{T}(\kappa z-\omega t)],
\end{equation}
with the artery's axis remaining fixed at a radial offset from the coordinate axis. Here, $\mathcal{T}(\kappa z-\omega t)$ is a (dimensionless) traveling wave profile characterized by its wavenumber $\kappa$, angular frequency $\omega$, and relative (dimensionless) amplitude $\epsilon$. Meanwhile, the changing outer radius, $r_e(z, t)$, is to be determined. The transverse dimension of the conduit is assumed to be much smaller than $\kappa^{-1}$. The axial length of the conduit is $\ell$. The eccentricity $ecc = d/r_0$ is the ratio of the radial offset between the axes of the two cylinders to the inner radius at equilibrium; $ecc=0$ corresponds to concentric cylinders, and $ecc = (r_{e,0} - r_0)/r_0$ corresponds to the artery making contact with the endfeet layer. Both the inner and outer walls are assumed to be impermeable to the flow. The porosity and permeability of the interior are denoted by $\phi$ and $\mathfrak{K}$, respectively. The fluid in the domain has density $\rho$ and dynamic viscosity $\mu$. 

\subsection{Lubrication approximation and axial velocity}

We denote dimensional and dimensionless variables, where possible, by lowercase and uppercase letters, respectively. Focusing on the key variables of the problem, we introduce the typical scalings for lubrication theory, i.e., the \emph{long-wave} approximation for the peristaltic wave \citep{Coenen2021}:
\begin{multline} 
    R=\frac{r}{r_0},\quad Z=\kappa z,\quad T=\omega t,\quad
    V_Z=\frac{\varv_z}{\epsilon \omega/\kappa},\quad 
    V_R=\frac{\varv_r}{\epsilon \omega r_0},\quad
    P= \frac{p-p_\mathrm{ref}}{\mu \epsilon \omega/(\kappa r_0)^2}.
    \label{eq: scalings}
\end{multline}

The annular flow is assumed to be at small Womersley number, $Wo=r_0\sqrt{\rho\omega/\mu}$, as is typical for flows in the PAS \citep{Boster2023}; note that $Wo^2 = \rho\omega r_0^2/\mu$ is the ratio of the unsteady inertial term to the transverse viscous term in the momentum equation. Under the scalings from \eqref{eq: scalings}, and making the assumptions that $Wo^2 \ll 1$ and $(\kappa r_0)^2 \ll1$ (see table~\ref{tab:parameter_table}), the inertial terms and axial viscous forces drop out of the momentum equation, leaving a balance of transverse viscous forces, the axial pressure gradient, and the drag from the enclosed porous medium. Further, since $P$ is not a function of $R$ or $\theta$, we can write $V_Z(R,\theta,Z)=-\left(\partial P/\partial Z \right)\hat{V}_Z(R,\theta)$. Then, $\hat{V}_Z(R,\theta)$ solves a boundary-value problem in the cross-section:
\begin{equation}
    \frac{1}{R} \frac{\partial}{\partial R} \left(R \frac{\partial \hat{V}_Z}{\partial R}\right)+\frac{1}{R^2} \frac{\partial^2 \hat{V}_Z}{\partial \theta^2} - \frac{1}{Da} \hat{V}_Z=-1, \qquad \hat{V}_Z=0 \;\;\text{on}\;\; \partial\Omega_\mathrm{cs},
\label{eq: crosssectional_velocity_eqn}
\end{equation}
where $Da=\mathfrak{K}/(\phi r_0^2)$ is the Darcy number, which quantifies the relative contribution of the porous medium drag \citep{Nield2017}. Here, $\partial\Omega_\mathrm{cs}$ is the boundary of the non-axisymmetric (eccentric annular) domain $\Omega_\mathrm{cs}$. For our problem, $\partial\Omega_\mathrm{cs}$ consists of two circles.  In the outer-wall-centered coordinate system of figure~\ref{fig:PVS_ug_schematic}, the inner part of $\partial\Omega_\mathrm{cs}$ is therefore the curve $R = R_{a,c} = ecc\,\cos\theta + \sqrt{R_a^2 - ecc^2\sin^2\theta}$ with $R_a = 1+\epsilon\mathcal{T}$, while the outer part of $\partial\Omega_\mathrm{cs}$ is $R=R_e$, where the form of $R_e$ will be discussed in \S\ref{sec:coupling}. For given (fixed) $R_a$ and $R_e$, we solved \eqref{eq: crosssectional_velocity_eqn} by the finite element method using FEniCS 2019.1.0, an open-source computing platform \citep{FEniCS1,FEniCS2}, as described in Appendix~\ref{app:monetum_eq_vv}.

Figure~\ref{fig:axial_velocity_profile} shows several solutions for the $\hat{V}_Z$ profile for different pairs of $ecc$ and $Da$ values. The profile develops localized boundary layers near the walls for $Da \ll 1$, taking on a plug-flow shape, and the velocity magnitude is proportional to $Da$; see \eqref{eq: crosssectional_velocity_eqn}.

Once we determine the cross-sectional profile, $V_Z$ (using $\hat{V}_Z$ from \eqref{eq: crosssectional_velocity_eqn}), the flow rate, $Q$, can be computed and related to the \emph{resistance}, $\mathcal{R}$, as follows:
\refstepcounter{equation}
$$
    \label{eq: flow rate resistance defn}
    Q = \iint_{\Omega_\mathrm{cs}} V_Z \, dA = - \frac{\partial P}{\partial Z} \frac{1}{\mathcal{R}},\qquad
    \mathcal{R} \defeq \left[\iint_{\Omega_\mathrm{cs}} \hat{V}_Z \, dA \right]^{-1},
\eqno{(\theequation{\mathit{a},\mathit{b}})}
$$
where $dA$ is the area element on annular domain $\Omega_{\mathrm{cs}}$.

\begin{figure}
    \centering
    \includegraphics[width=\textwidth]{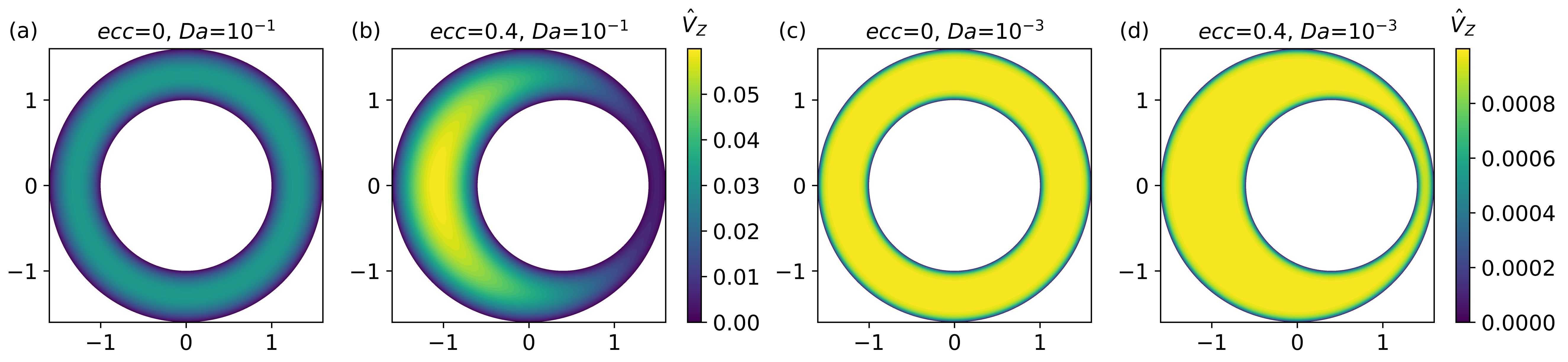}
    \caption{Contours of the axial velocity $\hat{V}_Z$ variation in the eccentric annular cross-section for (a,b) $Da=10^{-1}$ (weak porous drag, approaching an ``open space'') and (c,d) $Da=10^{-3}$ (representative of a porous medium), at two eccentricities each.}
    \label{fig:axial_velocity_profile}
\end{figure}

\subsection{Coupling flow and deformation}
\label{sec:coupling}

To couple the flow and deformation, we need to impose kinematic conditions at the inner (arterial wall) and outer (endfeet layer) radii. Under the long-wave approximation, for our chosen initially circular boundaries, these conditions reduce to matching the normal fluid velocity to the time-rate-of-change of the radius of each wall \citep{Zhang2024,Takagi2011}: $\bm{v}\bm{\cdot}\bm{n}|_a = \partial r_a/\partial t$ on the arterial (`$a$') wall and $\bm{v}\bm{\cdot}\bm{n}|_e = \partial r_e/\partial t$ at the endfeet (`$e$') layer, where in each case $\bm{n}$ is the outward normal of the respective circle (i.e., the radial direction measured from that cylinder's own axis). At the outer wall, the radius can expand due to the compliance of the endfeet layer, with $r_e(z,t) = r_{e,0} + u_e(z,t)$. We take the radial displacement to be directly proportional to the excess hydrodynamic pressure \citep{Takagi2024}, as $u_e(z,t) = \mathcal{C} \left[ p(z,t) - p_\mathrm{ref} \right]$, consistent with most fluid--structure interaction theories based on linear elasticity \citep{Christov2022,Rallabandi2024}. The proportionality constant is sometimes called the compliance, $\mathcal{C}$, and is determined by solving a suitable linear elasticity problem \citep{Christov2022}. Here, it is estimated (see table~\ref{tab:parameter_table}). 

Such a closure can be derived from the theory of elasticity, under the assumptions of (i) a long and slender condition (the long-wave approximation), which results in (ii) a state of plane strain in the elastic solid, decoupling axial cross-sections from each other, and (iii) tissue properties that are approximately homogeneous and isotropic. Under the lubrication approximation, the pressure in each cross-section is uniform to leading order in $(\kappa r_0)^2$. Thus, the outer circular wall of each cross-section is uniformly loaded by a normal stress (in the radial direction). Consequently, $u_e \propto p$ at each $z$. \citet{Wang2022FlowExperiments} provide an illustrated example of such a calculation for a similar configuration.

Now, recall that $R_{a,c} = ecc\,\cos\theta + \sqrt{R_a^2 - ecc^2\sin^2\theta}$ is the resolved location of the arterial wall in the given coordinate system, and the deformed radii are $R_a = 1 + \epsilon \mathcal{T}$ and $R_e = R_{e,0} +\epsilon \beta P$. Thus, for these circular inner and outer walls, introducing the dimensionless variables \eqref{eq: scalings}, the kinematic boundary conditions (BCs) are
\begin{subequations}\label{eq: v_wall_prime}
\begin{align}
    \label{eq: v_ra_prime}
    \bm{V}\bm{\cdot}\bm{n}|_a &= 
    \frac{1}{\epsilon}\frac{\partial R_{a,c}}{\partial T} = \frac{1}{\epsilon} \frac{R_a}{\sqrt{R_a^2 - ecc^2 \sin^2\theta}} \frac{\partial R_a}{\partial T} = -\frac{R_a}{\sqrt{R_a^2 - ecc^2 \sin^2\theta}} \mathcal{T}',\\
    \label{eq: v_re_prime}
    \bm{V}\bm{\cdot}\bm{n}|_e &=
    \frac{1}{\epsilon}\frac{\partial R_e}{\partial T} = \beta \frac{\partial P}{\partial T},
\end{align}\end{subequations}
where $\mathcal{T}'$ denotes derivative of $\mathcal{T}$ with respect to \emph{its argument}, $\Xi = Z-T$. In deriving \eqref{eq: v_re_prime}, we discover the dimensionless \emph{compliance number} $\beta = \mathcal{C} \mu \omega/(\kappa^2 r_0^3)$. Note that $\epsilon \beta$ is the ratio of the lubrication pressure scale, $\mu \epsilon \omega/(\kappa r_0)^2$, to the effective stiffness of the outer wall, $r_0/\mathcal{C}$.

\subsection{The hydraulic resistance}
\label{sec:resistance}

Extending the approach of \citet{Coenen2021}, we compute the resistance by domain perturbation. Assuming the relative wave amplitude of arterial pulsation is small ($\epsilon\ll1$), the total resistance can be written as a Taylor expansion:
\begin{equation}
    \mathcal{R} = \mathcal{R}_{0} \Bigg\{1 + \underbrace{\left[\frac{1}{\mathcal{R}}\frac{\partial \mathcal{R}}{\partial R_a}\right]_0}_{\Delta_1} \epsilon \mathcal{T} + \underbrace{\left[\frac{1}{\mathcal{R}}\frac{\partial \mathcal{R}}{\partial R_e}\right]_0}_{\Delta_{2}} \epsilon \beta P \Bigg\} + O(\epsilon^{2}),
    \label{eq: resistance_perturbation_eqn}
\end{equation}
where the zero subscripts denote the quantities evaluated on the \emph{undeformed} non-axi\-sym\-met\-ric annular space. The domain-perturbation approach requires that $\epsilon \mathcal{T} \ll 1$ and $\epsilon \beta P \ll R_{e,0}$, which we expect to hold when $\epsilon \ll 1$. We detail the numerical evaluation of the derivatives in the definitions of $\Delta_1$ and $\Delta_2$ in Appendix~\ref{app:Delta1_Delta2}. The perturbation expansion \eqref{eq: resistance_perturbation_eqn} is critical to the two-way coupling of peristalsis, flow, and outer-wall deformation, without having to solve \eqref{eq: crosssectional_velocity_eqn} and \eqref{eq: flow rate resistance defn} for every $Z$ and $T$. An alternative approach is the thin-gap approximation employed by \citet{Takagi2024}.

\begin{figure}
    \centering
    \includegraphics[width=\textwidth]{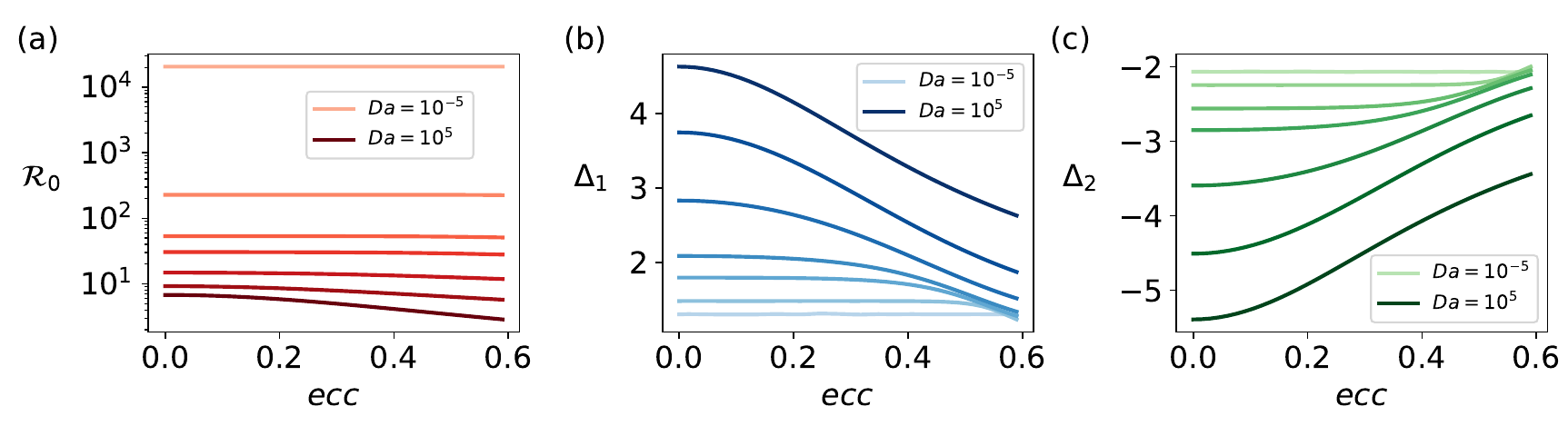}
    \caption{Computed dependence of the resistance perturbation expansion parameters, $\mathcal{R}_0$, $\Delta_1$, and $\Delta_2$ from \eqref{eq: resistance_perturbation_eqn}, on the  eccentricity, $ecc$, for selected values of the Darcy number, $Da = 10^{-5}, 10^{-3}, 5\times10^{-3}, 10^{-2}, 3\times10^{-2}, 10^{-1}$, and $10^5$ (light to dark curves).} 
    \label{fig:perturbation_expansion_parameters_vs_eccentricity}
\end{figure}

Figure~\ref{fig:perturbation_expansion_parameters_vs_eccentricity} shows $\mathcal{R}_0$, $\Delta_1$, and $\Delta_2$ as functions of the eccentricity, $ecc$, for various Darcy numbers across a wide range, $Da\in[10^{-5},10^{5}]$, to show all possible behaviors. First, from figure~\ref{fig:perturbation_expansion_parameters_vs_eccentricity}(a), we observe that the base resistance, $\mathcal{R}_0$, decreases by orders of magnitude with increasing $Da$, because the annular space is ``less'' resistive (more ``open''), and the viscous drag of the walls becomes dominant. Then, $\mathcal{R}_0$ decreases with $ecc$ because of the larger region away from the no-slip walls where the flow experiences less resistance (see figure~\ref{fig:axial_velocity_profile}(b)). Conversely, as $Da$ decreases, the viscous drag near the wall is confined to thin boundary layers  (figure~\ref{fig:axial_velocity_profile}), and the flow profile is plug-like throughout the domain \citep{Nield2017}, leading $\mathcal{R}_0$ to become largely independent of $ecc$ for $Da \ll 1$. This observation can also be rationalized via \eqref{eq: crosssectional_velocity_eqn} and \eqref{eq: flow rate resistance defn}, noting that for $Da\ll1$, the ``outer solution'' is $\hat{V}_Z \sim Da$ and thus $\mathcal{R}_0 \sim 1/(Da\, A)$, where $A$ is the area of the eccentric annular cross-section. For our problem, both boundaries are circles, and a standard exercise shows that $A = \pi(R_{e,0}^2 - 1) = 4.90$ is the dimensionless area of the annular space, which is independent of $ecc$ for our setup (figure~\ref{fig:PVS_ug_schematic}). Furthermore, using $\hat{V}_Z \sim Da$ and $\mathcal{R}_0 \sim 1/(Da\, A)$ for $Da\ll1$, it is easy to show that $\Delta_1 \sim 2\pi R_a/A \approx 1.28$ and $\Delta_2 \sim - 2\pi R_{e,0}/A \approx -2.05$, independent of $ecc$ as $Da\to0$, precisely matching the small-$Da$ curves' plateaus seen in figure~\ref{fig:perturbation_expansion_parameters_vs_eccentricity}(b,c), respectively.

Finally, from figure~\ref{fig:perturbation_expansion_parameters_vs_eccentricity}(b,c), we observe that $\Delta_1$ (due to arterial wave displacement) and $\Delta_2$ (due to endfeet-layer, i.e., outer-wall, displacement) exhibit similar trends with respect to $ecc$ and $Da$ when considering their magnitudes. However, while the arterial wave increases total resistance ($\Delta_1>0$) due to conduit constriction, brain compliance decreases it ($\Delta_2<0$) due to conduit expansion.

\subsection{The pressure}
\label{sec:pressure}

To derive the equation for axial pressure evolution that incorporates peristalsis at the inner wall and compliance at the outer wall of the fluidic domain, let us consider the dimensionless continuity equation. Integrating the dimensionless continuity equation over an arbitrary (potentially deformed) cross-section
$\Omega_\mathrm{cs}$, the axial term yields $\partial Q/\partial Z$ (using the Leibniz rule and the no-slip BCs on the walls), while the transverse divergence becomes a boundary flux by Gauss' theorem:
\begin{equation}
    \frac{\partial Q}{\partial Z}
    + \oint_{\partial\Omega_{\mathrm{cs}}}  \bm{V}\bm{\cdot}\bm{n} \, ds = 0,
    \label{eq: continuity_eqn_reduced}
\end{equation}
where $\bm{n}$ is the \emph{outward} normal to the fluid domain (i.e., it has opposite direction at the two walls) and $ds = rd\theta$ is an arclength element. Splitting up the line integral in \eqref{eq: continuity_eqn_reduced} into two parts, we obtain
\begin{equation}
    \frac{\partial Q}{\partial Z}
    - \int_0^{2\pi} \bm{V}\bm{\cdot}\bm{n}|_a \, R_{a,c} \, d\theta 
    + \int_0^{2\pi} \bm{V}\bm{\cdot}\bm{n}|_e \, R_e \, d\theta  
    = 0.
    \label{eq: continuity_eqn_reduced2_}
\end{equation}
Now, we use the kinematic BCs~\eqref{eq: v_wall_prime} to evaluate the integrals and eliminate $Q$ via (\ref{eq: flow rate resistance defn}\textit{a}). A straightforward but lengthy calculation shows that \eqref{eq: continuity_eqn_reduced2_} yields a
nonlinear PDE for the pressure:
\begin{equation}
    \underbrace{\frac{\partial}{\partial Z} \left(\frac{1}{\mathcal{R}} \frac{\partial P}{\partial Z} \right)}_{\text{viscous resistance}} - \underbrace{2 \pi \left(1+\epsilon \mathcal{T} \right) \mathcal{T}'}_{\text{peristalsis}}
    - \underbrace{2\pi \left(R_{e,0} + \epsilon \beta P\right) \beta \frac{\partial P}{\partial T}}_{\text{wall compliance}} = 0. 
    \label{eq: final_nonlinear_pressure_equation}
\end{equation}
The equation can be solved using a relevant set of BCs on $P$ at $Z=0,L$, including Dirichlet or resistance (Robin) ones. We focus on periodic BCs: $P(0,T)=P(L,T)$ for all $T>0$.

\begin{table}
    \small
    \centering
    \def~{\hphantom{0}}
    \begin{tabular}{lcccc}
        Parameter & Symbol & Range & Default & Unit\\[3pt]
        Artery radius & $r_0$ & 5--20 & 10 & \si{\micro\meter}\\
        PAS width & $r_{e,0}-r_0$ & 2--10 & 6 & \si{\micro\meter}\\
        PAS length\ & $\ell$ & -- & $2\lambda=2c/f=0.5$ & \si{\meter}\\
        Artery eccentricity & $ecc$ & 0--0.6 & 0--0.59 & --\\
        PAS porosity & $\phi$ & 0.5--0.9 & 0.8 & --\\
        PAS permeability & $\mathfrak{K}$ & $4.5 \times 10^{-15}$--$3.01 \times 10^{-12}$ & $8 \times 10^{-14}$--$8 \times 10^{-12}$ & \si{\meter\tothe{2}}\\
        Cardiac rel.\ wave ampl.\ & $\epsilon$ & 0.01--0.025 & 0.02 & --\\
        Cardiac wave frequency & $f$ & 2--6 & 4 & \si{\hertz}\\
        Cardiac wave speed & $c$ & 0.02--10 & 1 & \si{\meter\per\second}\\
        CSF viscosity & $\mu$ & $7 \times 10^{-4}$--$1 \times 10^{-3}$ & $1 \times 10^{-3}$ & \si{\pascal\second}\\
        CSF density & $\rho$ & 1000 & 1000 & \si{\kg\per\meter\tothe{3}}\\
        Brain compliance & $\mathcal{C}$ & $1 \times 10^{-11}$--$1 \times 10^{-8}$ & $2.5 \times 10^{-11}$--$2.5 \times 10^{-9}$ & \si{\meter\per\pascal}\\
        Long-wavelength parameter & $\kappa r_0$ & $6.28\times 10^{-6}$--$3.77\times 10^{-2}$ & $2.5\times 10^{-4}$ & --\\
        Womersley number & $Wo$ & $1.77\times 10^{-2}$--0.15 & $5\times 10^{-2}$ & --\\
        Darcy number & $Da$ & $5.62\times 10^{-5}$--$6.02\times 10^{-2}$ & $10^{-3}$--$10^{-1}$ & --\\
        Compliance number & $\beta$ & 0.398--397.9 & 1--100 & --\\    
    \end{tabular}
    \caption{Ranges of dimensional and dimensionless parameters for penetrating PAS flow based on  \citep{Bloomfield1998,Kedarasetti2020,Romano2020,Tithof2022,Boster2023,Gan2023}. The brain compliance is estimated as $\mathcal{C}\simeq {r_0}/{E_{\text{endfeet}}}$, where $E_{\text{endfeet}}\in[10^3, 10^6]$ is an ``elastic modulus'' in the cited works. The ranges for $Da$ and $\beta$ are computed by varying $\mathfrak{K}$ and $\mathcal{C}$, respectively, over their stated ranges while holding $r_0$, $\mu$, $f$, and $c$ at their default values, rounding the range edge points where appropriate. The $ecc$ range is obtained by varying the axes offset $d$ from $d=0$ (axisymmetric case, corresponding to $ecc=0$) to $d=r_{e,0}-r_0$ (contact with outer wall, corresponding to $ecc=0.6$, which is approached but not attained).}
    \label{tab:parameter_table}
\end{table}

We focus on periodic cardiac pulsations with $\mathcal{T}(\cdot)=\sin(\cdot)$. Let $P(Z,T)=\mathcal{P}(\Xi)$ be the pressure waveform in the moving frame $\Xi = Z-T$. In this case, a perturbation solution, $\mathcal{P} = \mathcal{P}_0 + \epsilon \mathcal{P}_1 + \cdots$ in $\epsilon\ll1$, of \eqref{eq: final_nonlinear_pressure_equation} can be constructed. After a lengthy but straightforward exercise (see Appendix~\ref{app:analytical_approx}), we find:
\refstepcounter{equation}
$$\label{eq: linearized P sol}
    \mathcal{P}_0(\Xi) = 
    \frac{2\pi\mathcal{R}_0}{1+B^2}
    + \frac{2\pi\mathcal{R}_0}{\sqrt{1+B^2}} \sin(\Xi+\phi_P),
    \qquad
    \phi_P = -\tan^{-1}(1/B),
\eqno{(\theequation{\mathit{a},\mathit{b}})}
$$
where we have introduced $B = 2 \pi \mathcal{R}_0 R_{e,0} \beta$ for convenience.

Simultaneously, we discretize \eqref{eq: final_nonlinear_pressure_equation} in time using the $\vartheta$-scheme \citep[see, e.g.,][Ch.~31]{FEniCS2} and in space using first-order Lagrange finite elements with a periodic domain constraint, as described in Appendix~\ref{app:pressure_eq_vv}. To shorten the transient to the periodic state, we used \eqref{eq: linearized P sol} to generate the initial condition $P(Z,0)$. The results are reported as the mean of the last three cycles. 

We verified our model's self-consistency by simulating the case $Da\to\infty$, $\beta=0$, and $ecc=0.2$, and confirming exact agreement of $Q(L/4,T)$ and $P(L/4,T)$ with predictions from an independent numerical implementation of equations (4.2) and (4.4) of \cite{Coenen2021} for this same geometry. Additional details on verification and validation, as well as the rationale for the choices of the mesh element count, the number of time steps per cardiac cycle, and the number of cardiac cycles, are provided in Appendix~\ref{app:pressure_eq_vv}.


\section{Results and discussion}
\label{sec:Results}

We now explore the implications of the outer wall's compliance in a porous annular conduit of length $\ell=2\lambda$ (thus, $L=\kappa\ell=4\pi$), consistent with the geometries considered by \citet{Carr2021} and \citet{Coenen2021}, and relative wave amplitude $\epsilon = 0.02$ (see table~\ref{tab:parameter_table} for default parameters) by simulating \eqref{eq: final_nonlinear_pressure_equation} subject to periodic BCs. The long conduit, while not representative of the typical shorter, bifurcation-free perivascular segments, yields a finite net pumping rate due to peristalsis, which we aim to understand in this work.

\subsection{Flow metrics for compliant conduits with different eccentricities}

\begin{figure}
    \centering
    \includegraphics[width=\textwidth]{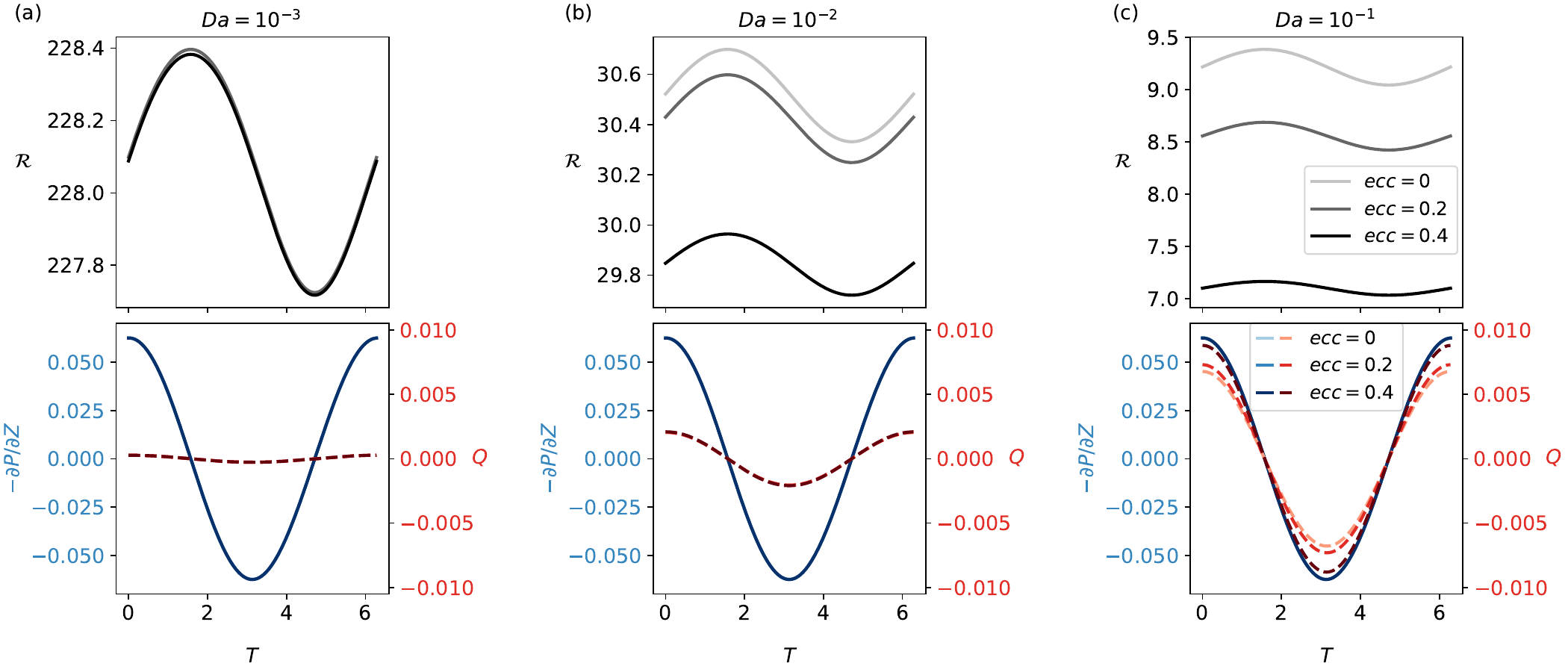}
    \caption{Effect of eccentricity on the flow metrics for (a) $Da=10^{-3}$ (note that all curves overlap), (b) $Da=10^{-2}$, and (c) $Da=10^{-1}$. Each column corresponds to a Darcy number, while the rows represent flow metrics, with resistance in the top row, and the pressure gradient and flow rate in the bottom row. All plots are at $Z=L/4$ and $\beta=10$.}
    \label{fig:ecctest}
\end{figure}

First, in figure~\ref{fig:ecctest}, we examine the effect of artery eccentricity, $ecc$, on the relevant flow metrics (resistance, pressure gradient, and flow rate). We consider three Darcy numbers ($Da=10^{-3}$, $10^{-2}$, and $10^{-1}$), corresponding to the annular conduit transitioning from a porous to an open space. In the top row of figure~\ref{fig:ecctest}, we observe that, for small $Da$, the resistance $\mathcal{R}$ is high, as expected, due to the substantial obstruction posed by the porous medium, and vice versa for larger $Da$. For $Da=10^{-3}$, the resistance curves nearly overlap for all $ecc$ values because the flow is dominated by the porous drag, and hence becomes independent of $ecc$ (recall figure~\ref{fig:perturbation_expansion_parameters_vs_eccentricity}). On the other hand, for the less porous spaces ($Da=10^{-2}$ and $10^{-1}$), resistance decreases with increasing eccentricity, and the curves are distinct.

Second, we examine the pressure gradient $-\partial P/\partial Z$ and the flow rate $Q$ in the bottom row of figure~\ref{fig:ecctest}. The pressure gradient is largely independent of $Da$ and $ecc$ because, for periodic BCs, it is driven by the peristaltic wave (the only forcing). Meanwhile, the resistance adjusts to the flow conditions, and the flow rate reflects it.  Specifically, the $Q$ amplitude increases with $Da$, as the space becomes more open. Likewise, the $Q$ amplitude increases with $ecc$ due to the emergence of a low-resistance channel away from the outer wall (recall figure~\ref{fig:axial_velocity_profile}(b)).

\subsection{Compliance of the outer wall and flow--pressure phasing}

\begin{figure}
    \centering
    \includegraphics[width=\textwidth]{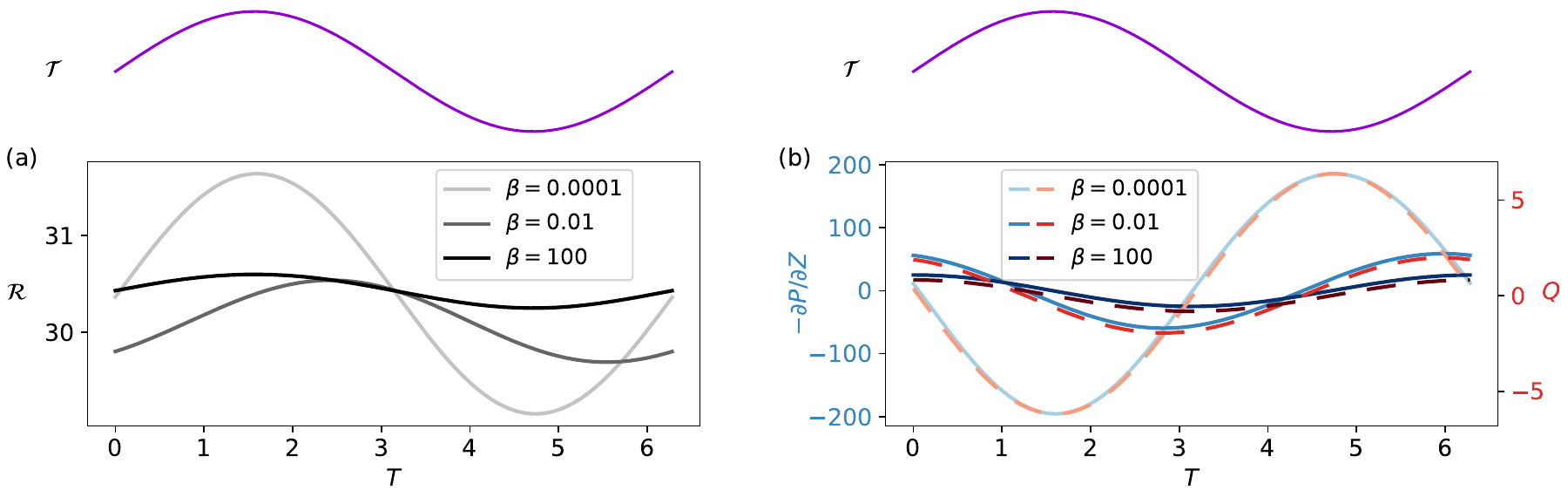}
    \caption{Effect of compliance on flow metrics. Time series of (a) resistance $\mathcal{R}$, (b) pressure gradient $-\partial P/\partial Z$ and flow rate $Q$. The driving peristaltic wave profile, $\mathcal{T}$, is shown above each plot. In (b), the $-\partial P/\partial Z$ and $Q$ curves corresponding to $\beta=100$ are multiplied by a factor of 4000 to highlight the phase relationship. All curves are for $Da=10^{-2}$, $ecc=0.2$, and $Z=L/4$.}
    \label{fig:betatest}
\end{figure}

Next, we study the effect of compliance across these scenarios by varying the compliance number in figure~\ref{fig:betatest}. Recall that, since $\beta = \mathcal{C} \mu \omega/(\kappa^2 r_0^3)$, the overall effect of compliance is not solely controlled by the elasticity-related constant $\mathcal{C}$ but also by the artery radius $r_0$, fluid viscosity $\mu$, wave frequency $\omega$, and wave number $\kappa$. In this figure, we only consider $Da=10^{-2}$; the trends are similar for other choices of $Da$. A smaller $\beta$ (stiffer outer wall) leads to higher amplitude of the resistance, pressure gradient, and flow rate, and vice versa. For a more compliant PAS, the outer wall expands in response to the hydrodynamic forces within, increasing the local cross-sectional area and reducing the amplitude of hydraulic resistance. The higher compliance also attenuates the pressure gradient amplitude by redistributing the driving energy into wall deformation. Consequently, the instantaneous flow rate amplitudes decrease with $\beta$.

In passing, we note that $\beta = 100$ means that $\epsilon\beta = 2 \not\ll 1$, yet this highly compliant case does not invalidate the domain-perturbation expansion introduced in \eqref{eq: resistance_perturbation_eqn} above because the actual approximation made can be rephrased as $\max |\epsilon\beta P| \ll R_{e,0}$. Upon using $P\simeq \mathcal{P}_0$ and \eqref{eq: linearized P sol}, we calculate
\begin{equation}
    \max |\epsilon\beta P| \simeq  \frac{\epsilon}{R_{e,0}} \max_B \frac{B(1+\sqrt{1+B^2})}{1+B^2} = \frac{3\sqrt{3} \epsilon}{4R_{e,0}} \approx 0.016 \ll 1.6 = R_{e,0}.
    \label{eq:max_e_b_P}
\end{equation}
This a posteriori check highlights that the wall displacement saturates: a more compliant wall deforms more readily but is loaded by a correspondingly smaller pressure.

The most prominent feature in figure~\ref{fig:betatest} is the \emph{phase shift} of $-\partial P/\partial Z$ and $Q$ (relative to the driving peristaltic wave $\mathcal{T}$), which is a strong function of $\beta$. The phase is evident from \eqref{eq: linearized P sol}. Specifically, we find that the phase of the pressure relative to the driving wave is $\phi_{P}$ given in (\ref{eq: linearized P sol}\textit{b}), with the pressure gradient shifted by a further quarter period, $\phi_{-\partial P/\partial Z} = \phi_P - \pi/2$ (see Appendix~\ref{app:periodic_bc}). For even moderate $\beta$, the phase saturates to $\phi_{-\partial P/\partial Z} \to - \pi/2$ rapidly since $\mathcal{R}_0$ is large; in the opposite limit $\phi_{-\partial P/\partial Z} \to -\pi$ as $\beta\to0$. Both of these limits are evident in figure~\ref{fig:betatest}(b), keeping in mind that a positive phase in $\Xi$ appears as a lag when plotting against $T$ at fixed $Z$ because $\Xi = Z - T$, and vice versa. 
Although previous work has noted that phase differences between the peristaltic wave, flow, pressure, and deformation can affect net glymphatic transport in brain tissue \citep{Li2025,Nozaleda2025}, no analytical connection was made to brain tissue compliance. Our model precisely quantifies this phase in terms of the key dimensionless parameters, specifically the compliance number.

\subsection{Net pumping rate}

\begin{figure}
    \centering
    \includegraphics[width=\textwidth]{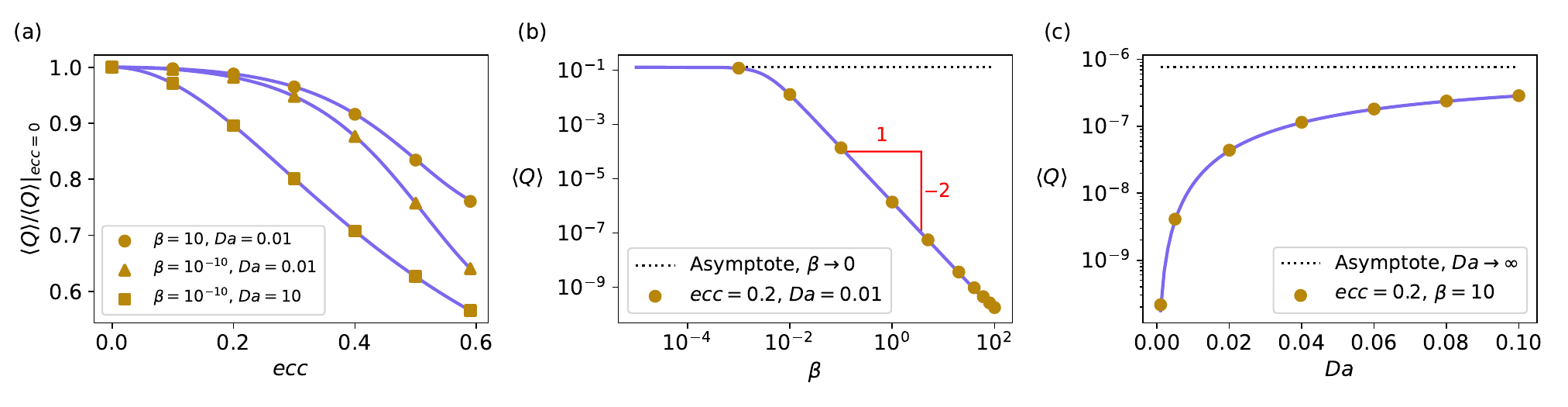}
    \caption{Pumping (space--time or cycle-averaged) flow rate as a function of (a) eccentricity, $ecc$, (b) compliance, $\beta$, and (c) Darcy number, $Da$. The darker (purple) curves are based on the analytical approximation \eqref{eq:analytical_pumping_rate} for $\epsilon\ll1$. Cases with $\beta = 10^{-10}$ represent the rigid limit evaluated through the present compliant model.} 
    \label{fig:SpaceTimeAvgQ}
\end{figure}

Finally, we discuss the net pumping (cycle-averaged) flow rate $\langle Q \rangle$. Figure~\ref{fig:SpaceTimeAvgQ}(a) shows that $\langle Q \rangle$ decreases with $ecc$, even for $Da<\infty$ and $\beta>0$. The case of $Da=10$ and $\beta=10^{-10}$ is illustrative of the rigid open channel of \citet{Coenen2021}. Meanwhile, our model can handle a wide range of $Da$ and $\beta$, as highlighted by the other two curves ($Da=0.01$ and $\beta=10^{-10}$ for a porous rigid channel, and $Da=0.01$ and $\beta=10$ for a porous compliant channel). 

Figure~\ref{fig:SpaceTimeAvgQ}(b) shows a new prediction of our model, namely that the net pumping is a strong function of compliance, decaying rapidly from the ``ceiling'' value (achieved at $\beta=0$ and $ecc=0$ for a given $Da$). A straightforward (but lengthy) perturbation calculation for $\epsilon \ll 1$ based on \eqref{eq: final_nonlinear_pressure_equation} (see Appendix~\ref{app:analytical_approx}) yields an analytical result:
\begin{equation}
    \langle Q \rangle
    = \frac{\pi \epsilon  \Delta_1}{1 + (2 \pi \mathcal{R}_0 R_{e,0}\beta)^2},
    \label{eq:analytical_pumping_rate}
\end{equation}
which rationalizes the observations of figure~\ref{fig:SpaceTimeAvgQ}(b). Two aspects of the result \eqref{eq:analytical_pumping_rate} are worth further interpretation. First, the compliance number always appears in the combination $B = 2\pi\mathcal{R}_0R_{e,0}\beta$, so that the \emph{onset} of the decay of $\langle Q \rangle$ is governed by the compliance number scaled by the base hydraulic resistance, and not by $\beta$ itself. Since $\mathcal{R}_0$ is large in a porous conduit (ranging from $\approx 30$ at $Da = 10^{-2}$ to $\approx 230$ at $Da = 10^{-3}$ in figure~\ref{fig:perturbation_expansion_parameters_vs_eccentricity}(a) at $ecc=0.2$) and $R_{e,0} = 1.6$, the ``crossover'' value $\beta^*$ is such that $1 \approx (2 \pi \mathcal{R}_0 R_{e,0}\beta^*)^2$ or $\beta^* \approx 3\times10^{-3}$ ($Da = 10^{-2}$)
and $4\times10^{-4}$ ($Da = 10^{-3}$). These compliance values are much smaller than the physiological range of table~\ref{tab:parameter_table}, implying that a penetrating PAS might always be in the $\beta^{-2}$ tail of \eqref{eq:analytical_pumping_rate}. In other words, even a very weakly compliant endfeet layer, which one might reasonably idealize as rigid, already suppresses the net pumping by orders of magnitude relative to the rigid-wall case. Second, $\mathcal{R}_0$ and $\Delta_1$ themselves depend on $ecc$ and $Da$ in a nontrivial manner (recall \S\ref{sec:methods}), so the crossover $\beta$ value changes with the geometry and the porosity.

Finally, figure~\ref{fig:SpaceTimeAvgQ}(c) shows that net pumping rapidly increases with $Da$, meaning that porous spaces strongly inhibit net pumping compared to open ones. Specifically, from our analytical approximation, we find that $\langle Q \rangle \sim \epsilon Da^2 A/(2 R_{e,0}^2 \beta^2)$ for $Da \ll 1$ from \eqref{eq:analytical_pumping_rate}, since $\mathcal{R}_0 \sim 1/Da$ and $\Delta_1 \sim 2\pi R_a/A$ as $Da \to 0$.

\section{Conclusion}
\label{sec:Conclusions}

Inspired by the mechanics of cerebrospinal fluid flow in the periarterial space, we developed a two-way-coupled fluid--structure interaction model for peristaltically driven flow in an eccentric annular porous conduit with a compliant outer wall. The slenderness of the conduit allows the use of a long-wave (lubrication) approximation, reducing the problem to an unsteady PDE for the pressure (after pre-solving an 
equation for cross-sectional velocity). We calculated changes in hydraulic resistance due to peristalsis (arterial displacements) and deformation of the surrounding brain tissue using a perturbation expansion in the peristaltic wave amplitude.

The resulting computationally (and analytically) tractable model for flow in \emph{non-axisymmetric} spaces with a \emph{compliant} outer wall in hand allowed us to perform parameter-space studies, focusing on the effect of the eccentricity of space, the Darcy number (quantifying the porous drag within the space), and the compliance number (elasticity of the space's outer confining wall) on the hydraulic resistance, the pressure gradient, and the flow rate. In contrast to previous studies on rigid conduits \citep{Carr2021,Coenen2021}, we discovered that compliance strongly reduces the net pumping rate, consistent with the simulations of \citet{Gan2023}.
Analytical approximations of our reduced model under periodic boundary conditions revealed a simple formula showing that the net pumping rate decays as the inverse square of the compliance number. Crucially, the onset of this decay is set by the compliance number scaled by the base resistance, which is large in a porous space; consequently, pumping is strongly reduced even for walls that are only weakly compliant. Similarly, we found that net pumping decreases in more obstructed spaces (i.e., at smaller Darcy numbers). 

A future direction could be to consider other boundary conditions, such as upstream and downstream resistance boundary conditions, which are typically encountered when modeling a finite PAS segment. We could also extend the model to account for a porous \emph{outer} wall \citep{Gan2023,Takagi2024}, which would require updating the boundary condition at the outer wall and incorporating the transverse variation of pressure and velocity at $O((\kappa r_0)^2)$ \citep{Coenen2021}. The transverse pressure variations could result in non-uniform deformation of the PAS outer wall for non-zero eccentricities.
Other driving mechanisms of flow in the PAS can be considered, such as functional hyperemia, which has a non-zero mean wave profile, unlike the sinusoidal wave profile used in our analysis. 
Finally, when the Womersley number is not small, it would be relevant to understand how \emph{elastoinertial rectification} \citep{Zhang2024}  contributes to the net pumping in non-axisymmetric geometries. It will also be relevant to analyze the role of the complex rheology of biofluids \citep{Covarrubias2026} in non-axisymmetric geometries.


\begin{bmhead}[Acknowledgements]
Simulations were performed using the community clusters of the Rosen Center for Advanced Computing at Purdue University. We thank Juan Andr\'{e}s Gil Duque for helpful explorations of the model in its earlier stages.
\end{bmhead}

\begin{bmhead}[Funding]
This research was supported by the National Aeronautics and Space Administration (NASA) under FINESST Research Grant No.\ 80NSSC25K7350.
\end{bmhead}

\begin{bmhead}[Declaration of interests]
The authors report no conflict of interest.
\end{bmhead}

\begin{bmhead}[Data availability statement]
The scripts to reproduce the data that support the findings of this study are openly available in the Purdue University Research Repository (PURR) at \url{https://doi.org/10.4231/NJ7B-5N87}.
\end{bmhead}

\begin{bmhead}[Author contributions]
\textbf{Nishanth Surianarayanan:} Conceptualization, Formal analysis, Data curation, Software, Validation, Investigation, Visualization, Methodology, Writing -- original draft, Writing -- review \& editing, Funding acquisition. \textbf{Ivan C. Christov:} Conceptualization, Formal analysis, Supervision, Validation, Investigation, Methodology, Writing -- original draft, Writing -- review \& editing, Project administration, Funding acquisition.
\end{bmhead}






\begin{appen}

\section{Detailed derivation of the analytical approximation(s) for $\epsilon \ll 1$}
\label{app:analytical_approx}

\subsection{Transformation to a moving frame}

Introducing the traveling wave pressure form $\mathcal{P}(\Xi = Z-T) = P(Z,T)$ so that $\partial P/\partial T = -\mathcal{P}'$, $\partial P/\partial Z = +\mathcal{P}'$, and $(\,\cdot\,)' = d/d\Xi$, we transform \eqref{eq: final_nonlinear_pressure_equation} into 
\begin{equation}
    \left( \frac{1}{\mathcal{R}} \mathcal{P}' \right)' - 2 \pi (1 + \epsilon \mathcal{T}) \mathcal{T}'
     + 2\pi \left( R_{e,0} + \epsilon\beta\mathcal{P} \right) \beta \mathcal{P}' = 0. 
    \label{eq: nonlinear_pressure_equation_periodic}
\end{equation} 

To obtain the pumping rate, we integrate \eqref{eq: nonlinear_pressure_equation_periodic} with respect to $\Xi$:
\begin{equation}
    \frac{1}{\mathcal{R}} \mathcal{P}'
    - 2 \pi \left( \mathcal{T} + \epsilon \mathcal{T}^2/2 \right)
     + \beta \left( 2\pi R_{e,0} \mathcal{P} + 2\pi\epsilon\beta\mathcal{P}^2/2 \right) + C_\lambda = 0,
    \label{eq: 1st int periodic_p}
\end{equation}
where $C_\lambda$ is determined by the constraint that $\langle \mathcal{P}'\rangle \equiv \delta P_\lambda/(2\pi)$, which represents some (arbitrary) pressure change per wavelength. To impose this constraint, we multiply \eqref{eq: 1st int periodic_p} by $\mathcal{R}$ and average over a cycle, i.e., we apply $\langle \, \cdot \, \rangle = \frac{1}{2\pi} \int_0^{2\pi} ( \, \cdot \, ) \, d\Xi$, to find
\begin{equation}
    C_\lambda = \frac{2 \pi \Big( \langle\mathcal{T}\mathcal{R}\rangle + \epsilon \langle\mathcal{T}^2\mathcal{R}\rangle/2 \Big)
     - \beta \Big( 2\pi R_{e,0} \langle\mathcal{P} \mathcal{R}\rangle + 2\pi\epsilon\beta\langle\mathcal{P}^2\mathcal{R}\rangle/2 \Big) - \delta P_\lambda/(2\pi)}{\langle \mathcal{R} \rangle }.
    \label{eq: C_lam periodic_p}
\end{equation}
Now, returning to \eqref{eq: 1st int periodic_p} and recognizing that the first term on the left-hand side is $-Q$, we once again average and solve for 
\begin{equation}
    \langle Q \rangle = C_\lambda 
    - 2\pi \left(\langle \mathcal{T} \rangle - \beta R_{e,0}\langle \mathcal{P} \rangle\right)
    - \pi \epsilon \Big(  \langle \mathcal{T}^2 \rangle - \beta^2 \langle\mathcal{P}^2 \rangle \Big) .
    \label{eq: periodic_pumping_rate}
\end{equation}

Note that this result is not closed; both $\mathcal{R}$ and $\mathcal{P}$ still need to be computed consistently from the coupled fluid--structure interaction problem. 
To overcome this difficulty and obtain some analytical insight, we proceed by a perturbation expansion in $\epsilon \ll 1$.

\subsection{Leading-order pressure solution in the moving frame}

Performing a perturbation expansion $\mathcal{P} = \mathcal{P}_0 + \epsilon \mathcal{P}_1 + O(\epsilon^2)$, consistent with the perturbation expansion for $\mathcal{R}$ in \eqref{eq: resistance_perturbation_eqn}, and substituting it into \eqref
{eq: nonlinear_pressure_equation_periodic}, we obtain a linear ordinary differential equation for $\mathcal{P}_0$:
\begin{equation}
    \frac{1}{\mathcal{R}_0} \mathcal{P}_0'' - 2 \pi \mathcal{T}'
    + 2\pi R_{e,0} \beta \mathcal{P}_0' = 0. 
    \label{eq: nonlinear_pressure_equation_periodic_linearized}
\end{equation}
This equation captures the \emph{leading-order} contributions of both peristalsis \emph{and} compliance.

Multiplying \eqref{eq: nonlinear_pressure_equation_periodic_linearized} by $\mathcal{R}_0$ and integrating once, we have
\begin{equation}
    \mathcal{P}_0' - 2 \pi \mathcal{R}_0 \mathcal{T} 
    + B \mathcal{P}_0 = B C_1, \qquad B = 2 \pi \mathcal{R}_0 R_{e,0} \beta . 
\end{equation}
Using the integrating factor $e^{B\Xi}$, we can solve the latter:
\begin{equation}
    \mathcal{P}_0(\Xi) = 2 \pi \mathcal{R}_0 e^{-B\Xi} \int \mathcal{T}(\hat{\Xi}) e^{B\hat{\Xi}} \, d\hat{\Xi} + C_1 + C_2 e^{-B\Xi}. 
    \label{eq: linearized travel P solution gen}
\end{equation}
Thus, we have obtained leading-order analytical solution $\mathcal{P}_0$, for the pressure variation, in terms of the base resistance $\mathcal{R}_0$, the compliance number $\beta$, and the average outer radius $R_{e,0}$, for a given peristaltic waveform $\mathcal{T}$. 
For $\mathcal{T}(\Xi) = \sin(\Xi)$, \eqref{eq: linearized travel P solution gen} yields:
\begin{equation}
    \mathcal{P}_0(\Xi) = 2 \pi \mathcal{R}_0 \left( \frac{B \sin \Xi - \cos \Xi}{1 + B^2} \right) + C_1 + \; C_2 e^{-B\Xi}. 
    \label{eq: linearized travel P solution}
\end{equation}

\subsection{Periodic BCs}
\label{app:periodic_bc}

For periodic BCs, we enforce $\mathcal{P}(0) = \mathcal{P}(2\pi)$ on the solution from \eqref{eq: linearized travel P solution}, to obtain
\refstepcounter{equation}
$$
    C_2 = 0,\qquad C_1 = \text{arbitrary}.
\eqno{(\theequation{\mathit{a},\mathit{b}})}
$$
Without loss of generality, $C_1$ is chosen by requiring that $\mathcal{P}_0(0)=0$, thus:
\begin{equation}
    \mathcal{P}_0(\Xi) = 2 \pi \mathcal{R}_0 \left( \frac{1 - \cos \Xi + B \sin \Xi}{1 + B^2} \right). 
    \label{eq: linearized travel P solution gauge}    
\end{equation}
We verify in \S\ref{app:pumping_rate} below that the net pumping rate is
independent of the choice $\mathcal{P}_0(0)=0$.

Equivalently, in amplitude--phase form, \eqref{eq: linearized travel P solution gauge} becomes
\begin{equation}
    \mathcal{P}_0(\Xi) = \frac{2\pi\mathcal{R}_0}{1+B^2}
    + \frac{2\pi\mathcal{R}_0}{\sqrt{1+B^2}} \sin(\Xi+\phi_P),
    \qquad \phi_P = -\tan^{-1}(1/B),
    \label{eq: P0 amplitude phase}
\end{equation}
taking the principal branch of $\tan^{-1}$ since  $\cos\phi_P = B/\sqrt{1+B^2}\ge0$ and $\sin\phi_P = -1/\sqrt{1+B^2}<0$; thus, $\phi_P\in[-\pi/2,0)$. Differentiating \eqref{eq: P0 amplitude phase}, and using $-\cos \xi = \sin(\xi - \pi/2)$,
\begin{equation}
    -\frac{\partial P}{\partial Z} = -\mathcal{P}_0'
    = \frac{2\pi\mathcal{R}_0}{\sqrt{1+B^2}}\,\sin\!\left(\Xi+\phi_P-{\pi}/{2}\right),
    \label{eq: dPdZ amplitude phase}
\end{equation}
so that $\phi_{-\partial P/\partial Z} = \phi_P - \pi/2$, with $\phi_{-\partial P/\partial Z}\to-\pi$ as $\beta\to0$ and $\to-\pi/2$ as $\beta\to\infty$. 
The rigid limit provides a sanity check: setting $B=0$ in \eqref{eq: linearized travel P solution gauge} gives $-\partial P/\partial Z = -2\pi\mathcal{R}_0\sin\Xi$ (anti-phase with the peristaltic wave).

\subsection{Pumping rate}
\label{app:pumping_rate}

To proceed, we continue with the $\epsilon \ll 1$ perturbation expansion by substituting \eqref{eq: resistance_perturbation_eqn} into \eqref{eq: periodic_pumping_rate}, having eliminated $C_\lambda$ using \eqref{eq: C_lam periodic_p}, to obtain
\begin{multline}
    \langle Q \rangle = \left[2 \pi \Big( \langle\mathcal{T}{\mathcal{R}_0}\rangle + \epsilon \Delta_1 \langle\mathcal{T}^2 {\mathcal{R}_0}\rangle + \epsilon \beta \Delta_2 \langle\mathcal{T}\mathcal{P}{\mathcal{R}_0}\rangle + \epsilon \langle\mathcal{T}^2{\mathcal{R}_0}\rangle/2 \Big) + O(\epsilon^2) \right]\\
    \times \frac{1}{\langle {\mathcal{R}_0} \rangle} \left[1-\epsilon\big(\Delta_1\langle\mathcal{T} \rangle+\Delta_2\beta \langle\mathcal{P}\rangle\big) + O(\epsilon^2) \right]\\
     - \left[\beta \Big( 2\pi R_{e,0} \langle\mathcal{P} {\mathcal{R}_0}\rangle + 2\pi R_{e,0} \epsilon \Delta_1 \langle\mathcal{P}\mathcal{T} {\mathcal{R}_0}\rangle + 2\pi R_{e,0} \epsilon\beta \Delta_2 \langle\mathcal{P}^2{\mathcal{R}_0}\rangle + 2\pi\epsilon\beta\langle\mathcal{P}^2{\mathcal{R}_0}\rangle/2 + O(\epsilon^2) \Big)\right]\\
    \times \frac{1}{\langle {\mathcal{R}_0} \rangle} \left[1-\epsilon\big(\Delta_1\langle\mathcal{T} \rangle+\Delta_2\beta \langle\mathcal{P}\rangle\big) + O(\epsilon^2) \right]\\
     - \pi \epsilon \Big(  \langle \mathcal{T}^2 \rangle  -\beta^2 \langle\mathcal{P}^2 \rangle \Big) - 2\pi \Big(\langle \mathcal{T} \rangle - \beta R_{e,0}\langle \mathcal{P} \rangle\Big) + O(\epsilon^2).
\end{multline}
Note that $\mathcal{R}_0$ is independent of $\Xi$, thus $\langle \mathcal{R}_0\rangle = \mathcal{R}_0$ and this term was factored out and canceled.
We have also carried along the contribution from the Taylor series expansions of the denominators. Regrouping terms, we find many self-cancel and $\langle\mathcal{T}\rangle=0$, due to the periodicity of the peristaltic wave, removes several others. Substituting the perturbation expansion for $\mathcal{P}$, we obtain
\begin{equation}
    \langle Q \rangle = 2 \pi \epsilon \Big[ \Delta_1 \langle\mathcal{T}^2\rangle + \beta ( \Delta_2 - R_{e,0} \Delta_1 )\langle\mathcal{T}\mathcal{P}_0\rangle
     -  \beta^2 R_{e,0} \Delta_2 \big( \langle\mathcal{P}_0^2\rangle - \langle\mathcal{P}_0\rangle^2 \big) \Big] + O(\epsilon^2) .
    \label{eq: periodic_pumping_rate_perturbation}
\end{equation}
We observe that compliance introduces two new terms: one proportional to the cross-correlation between the peristaltic and pressure waveforms, and the other proportional to the variance of the pressure waveform. Setting $\beta=0$, we recover equation (4.5) of \citet{Coenen2021}.

Note that setting $\mathcal{P}_0(0)=0$ leads to $\langle \mathcal{P}_0\rangle \ne 0$. While the physical gauge is fixed by the reference pressure $p_\mathrm{ref}$ in \S\ref{sec:methods}, fixing $\mathcal{P}_0(0)=0$ means that the solution \eqref{eq: linearized travel P solution gauge} corresponds to placing the pressure reference at $\Xi = 0$. Importantly, two features of the problem render our results insensitive to the arbitrary choice $\mathcal{P}_0(0)=0$. First, the leading-order equation \eqref{eq: nonlinear_pressure_equation_periodic_linearized} involves $\mathcal{R}_0$ only, hence $\mathcal{P}_0$ is only determined up to the additive constant, $C_1$. Second, and more importantly, the net pumping rate in \eqref{eq: periodic_pumping_rate_perturbation} is invariant under $\mathcal{P}_0 \mapsto \mathcal{P}_0 + C_1$ because, even though $\langle\mathcal{T}\mathcal{P}_0\rangle
\mapsto \langle\mathcal{T}\mathcal{P}_0\rangle + C_1 \langle\mathcal{T}\rangle$, $C_1 \langle\mathcal{T}\rangle = 0$ by periodicity of the peristaltic wave. Meanwhile the term $\langle\mathcal{P}_0^2\rangle - \langle\mathcal{P}_0\rangle^2$ in \eqref{eq: periodic_pumping_rate_perturbation} being a variance is unaffected by constant shifts.

Finally, $\langle\mathcal{P}_0\rangle^2$, $\langle\mathcal{P}_0^2\rangle$, and $\langle \mathcal{T} \mathcal{P}_0 \rangle$ in \eqref{eq: periodic_pumping_rate_perturbation} are evaluated based on the leading-order solution~\eqref{eq: linearized travel P solution} with $C_2=0$, $C_1 = 2\pi\mathcal{R}_0/(1+B^2)$, and $\mathcal{T}(\Xi) = \sin(\Xi)$. Specifically, we compute:
\begin{subequations}\label{eq: TP avg}\begin{alignat}{3}
    \langle \mathcal{P}_0 \rangle^2 &= \left[\frac{1}{2\pi} \int_0^{2\pi} \mathcal{P}_0(
    \Xi) \, d\Xi\right]^2 &&= 4 \pi^{2} \mathcal{R}_{0}^{2} \frac{1}{(1 + B^{2})^{2}},\\
    \langle \mathcal{P}_0^2 \rangle &= \frac{1}{2\pi} \int_0^{2\pi} \big[\mathcal{P}_0(
    \Xi)\big]^2 \, d\Xi &&= 2\pi^2 \mathcal{R}_0^2 \frac{3 + B^2}{(1 + B^2)^2}, 
    \label{eq: P square avg}\\
    \langle \mathcal{T} \mathcal{P}_0 \rangle &= \frac{1}{2\pi} \int_0^{2\pi} \mathcal{T}(\Xi)\mathcal{P}_0(
    \Xi) \, d\Xi &&= \pi \mathcal{R}_0 \frac{B}{1 + B^2}. 
\end{alignat}\end{subequations}
Using \eqref{eq: TP avg} in \eqref{eq: periodic_pumping_rate_perturbation}, we can  evaluate the net pumping rate $\langle Q \rangle$ for $\epsilon\ll1$:
\begin{equation}
    \langle Q \rangle 
    = \frac{\pi \epsilon  \Delta_1}{1 + B^2}
    \sim
    \begin{cases}
    \pi \epsilon \Delta_1, &\quad \beta \to 0,\\
    \pi \epsilon \Delta_1 \left( \frac{1}{4 \pi^2 \mathcal{R}_0^2 R_{e,0}^2} \right) \beta^{-2}, &\quad \beta \to \infty.
    \end{cases}
\end{equation}

\section{Finite element method simulation details: weak forms, verification, and validation}
\label{app:numerical_details}

\subsection{The axial momentum equation}
\label{app:monetum_eq_vv}

\subsubsection{Weak form}
The axial momentum equation~\eqref{eq: crosssectional_velocity_eqn} can be written in a weak (variational) form by standard steps \citep{FEniCS1,FEniCS2}. The variational problem to be solved is then: find $\hat{V}_Z$ such that, for every $\Psi$, $a(\hat{V}_Z,\Psi)=l(\Psi)$, where the bilinear and linear forms are, respectively, 
\begin{subequations}\begin{align}
    a(\hat{V}_Z,\Psi) &= \int_{\Omega_{\mathrm{cs}}} \left[\nabla \hat{V}_Z \cdot \nabla \Psi + \frac{1}{Da} \hat{V}_Z \Psi\right] dA,\\
    l(\Psi) &= \int_{\Omega_{\mathrm{cs}}} \Psi \, dA.
\end{align}\label{eq:final_axial_weak_form}\end{subequations}
Here, $\Psi$ is a `test' function from a first-order Lagrange finite element space \citep{FEniCS2}. As before, $\Omega_{\mathrm{cs}}$ is the domain representing the cross-section of the conduit, and $dA$ is the area element. The no-slip BC $\hat{V}_Z = 0$ on $\partial \Omega_\mathrm{cs}$ was strongly imposed using the FEniCS' built-in \texttt{DirichletBC}, eliminating those degrees of freedom from the subsequent matrix problem.

The two-dimensional computational geometry of the eccentric annular cross-section was constructed and meshed using the mesh-generation tool \texttt{mshr}. Then, the solution \eqref{eq:final_axial_weak_form} is obtained using the \texttt{solve} function in FEniCS with its default settings, which internally calls \texttt{LinearVariationalSolver}.

\subsubsection{Verification and validation}
\label{subsubsection: axial_mom_v_v}

We performed a verification of the axial momentum equation's solution by computing the relative errors for the velocity field and hydraulic resistance values for meshes with $216\,532$, $863\,743$, and $3\,459\,071$ elements (corresponding to \texttt{mshr} refinement level parameters of $300$, $600$, and $1200$, respectively at $ecc=0.2$). The verification was performed for the geometry with $ecc=0.2$ and $Da=10^{-5}$. The velocity $L^2$ error decreased from $2.17\%$ at $216\,532$ elements to $0.60\%$ at $863\,743$ elements, relative to the fine-mesh reference solution, which translates to a corresponding decrease in resistance error from $0.19\%$ to $0.04\%$ over the same mesh refinement. This monotonic decrease in both error metrics with increasing mesh resolution confirms convergence of the numerical solution, and the small resistance error (under $0.2\%$ even at the coarsest mesh) indicates that the mesh with $863\,743$ elements (refinement level $600$) provides sufficient accuracy for resistance calculations in this geometry.

To validate the numerical implementation, we compared the finite element solution against an analytical solution available for the concentric annulus case $(ecc = 0)$. In this case, since the domain is an axisymmetric annulus, the analytical solution of the boundary-value problem~\eqref{eq: crosssectional_velocity_eqn} is easily found in terms of modified Bessel functions of the first and second kinds ($\mathrm{I}_0$ and $\mathrm{K}_0$):
\begin{multline}
    \hat{V}_Z(R) = Da 
    - \frac{Da\left[\mathrm{K}_0\!\left(\dfrac{R_e}{\sqrt{Da}}\right) - \mathrm{K}_0\!\left(\dfrac{R_a}{\sqrt{Da}}\right)\right]}
    {\mathrm{I}_0\!\left(\dfrac{R_a}{\sqrt{Da}}\right)\mathrm{K}_0\!\left(\dfrac{R_e}{\sqrt{Da}}\right) - \mathrm{I}_0\!\left(\dfrac{R_e}{\sqrt{Da}}\right)\mathrm{K}_0\!\left(\dfrac{R_a}{\sqrt{Da}}\right)}\,
    \mathrm{I}_0\!\left(\frac{R}{\sqrt{Da}}\right)\\
    - \frac{Da\left[\mathrm{I}_0\!\left(\dfrac{R_a}{\sqrt{Da}}\right) - \mathrm{I}_0\!\left(\dfrac{R_e}{\sqrt{Da}}\right)\right]}
    {\mathrm{I}_0\!\left(\dfrac{R_a}{\sqrt{Da}}\right)\mathrm{K}_0\!\left(\dfrac{R_e}{\sqrt{Da}}\right) - \mathrm{I}_0\!\left(\dfrac{R_e}{\sqrt{Da}}\right)\mathrm{K}_0\!\left(\dfrac{R_a}{\sqrt{Da}}\right)}\,
    \mathrm{K}_0\!\left(\frac{R}{\sqrt{Da}}\right).
\label{eq:V_Z_ecc=0}
\end{multline}

The numerical solution on the default grid was compared with the analytical profile~\eqref{eq:V_Z_ecc=0} for the same concentric (axisymmetric) annular geometry and $Da=10^{-5}$ (a ``worst case scenario'' to ensure proper resolution of the boundary layer near the outer wall for $Da \ll 1$). We computed the relative $L^2$ error
\begin{equation}
    \varepsilon_{L^2} = \frac{\left\lVert \hat{V}_{Z,\mathrm{analytical}} - \hat{V}_{Z,\mathrm{numerical}} \right\rVert_{L^2(\Omega_\mathrm{cs})}}{\left\lVert \hat{V}_{Z,\mathrm{analytical}} \right\rVert_{L^2(\Omega_\mathrm{cs})}}
\end{equation}
over the full two-dimensional annular domain $\Omega_\mathrm{cs}$ using built-in FEniCS functions. For the mesh with $864\,834$ elements (corresponding to refinement level $600$ at $ecc=0$; note that the unstructured mesh generation leads to slightly different element counts for different $ecc$ values), the relative $L^2$ error was $0.7\%$, indicating strong agreement between the numerical and analytical solutions.

\subsection{Resistance contributions}
\label{app:Delta1_Delta2}

The derivatives in \eqref{eq: resistance_perturbation_eqn} were calculated by finite differences, using the information post-processed from the $\hat{V}_Z$ solution. Specifically, a central-difference scheme was used as follows:
\begin{subequations}\begin{align}
    \Delta_1 &= \left[\frac{1}{\mathcal{R}}\frac{\partial \mathcal{R}}{\partial R_a}\right]_0 = \frac{1}{\mathcal{R}_0} \left[\frac{\mathcal{R}|_{R_a = 1+h}-\mathcal{R}|_{R_a=1-h}}{2h}\right]_{\text{fixed $R_e$}},\\
    \Delta_{2} &= \left[\frac{1}{\mathcal{R}}\frac{\partial \mathcal{R}}{\partial R_e}\right]_0 = \frac{1}{\mathcal{R}_0} \left[\frac{\mathcal{R}|_{R_e=R_{e,0}+h}-\mathcal{R}|_{R_e=R_{e,0}-h}}{2h}\right]_{\text{fixed $R_a$}}.
\end{align}\label{eq: central_difference_scheme_Delta}\end{subequations}
Note that these derivatives are evaluated at fixed axes offset $d$, and therefore at fixed $ecc$ by construction.

In the FEniCS code, we calculated the resistances appearing in  \eqref{eq: central_difference_scheme_Delta} by perturbing the boundary and regenerating the mesh, with either the inner or outer circle fixed (as denoted). Then, we evaluated $\mathcal{R}$ according to (\ref{eq: flow rate resistance defn}\textit{b}) on the new geometry. We limited the eccentricity to $ecc \leq 0.59$ to avoid the degenerate/tangent case (here, $ecc = 0.6$), for which evaluating $\mathcal{R}|_{R_a = 1+h}$ or $\mathcal{R}|_{R_e = R_{e,0} - h}$ in \eqref{eq: central_difference_scheme_Delta} is not possible due to the intersection of the circles.

Next, we performed a convergence study to select the spacing $h$ used in the central-difference computation of $\Delta_1$ and $\Delta_2$, for the geometry chosen and the cross-sectional mesh size determined in Appendix~\ref{subsubsection: axial_mom_v_v}, and we verified the numerical calculation for representative Darcy numbers spanning the range considered. We tested a range of $h$ values and identified $h = 10^{-3}$, $5\times10^{-4}$, and $2.5\times10^{-4}$ as leading to suitably converged values of $\Delta_1$ and $\Delta_2$ (less than 1\% change). We selected $h = 5\times10^{-4}$ for subsequent computations. As an additional validation, we calculated the value of $\Delta_1$ in the limiting case $Da \to \infty$, and compared it to the values shown in figure 2 of \cite{Coenen2021}, finding agreement within approximately 1\%.

\subsection{The reduced pressure equation}
\label{app:pressure_eq_vv}

\subsubsection{Weak form}

The pressure PDE \eqref{eq: final_nonlinear_pressure_equation} derived above can be recast in the form
\begin{equation}
    f(P) + g(P) \frac{\partial P}{\partial T} = 0. 
    \label{eq: ex_PDE_theta_scheme}
\end{equation}
Now, we can apply the so-called $\vartheta$-scheme \citep[see, e.g.,][Ch.~31]{FEniCS2} to time-march \eqref{eq: ex_PDE_theta_scheme} as follows:
\begin{equation}
    \left[\vartheta f \left(P^{n} \right)+\left(1-\vartheta \right) f(P^{n+1}) \right]+\left[\vartheta g \left(P^{n} \right)+ \left(1-\vartheta \right) g(P^{n+1}) \right] \frac{P^{n+1} - P^{n}}{\Delta T}=0.
    \label{eq: theta_scheme}
\end{equation}
In this convention, $\vartheta$ weights the \emph{previous} time level, so that $\vartheta=0$ is the backward (implicit) Euler scheme, $\vartheta=1/2$ is the Crank--Nicolson scheme, and $\vartheta=1$ is the forward (explicit) Euler scheme. Specifically, on substituting the implied $f$ and $g$ functions, \eqref{eq: theta_scheme} becomes
\begin{multline}
    \vartheta \left[\frac{\partial}{\partial Z} \left(\frac{1}{\mathcal{R}^n} \frac{\partial P^n}{\partial Z}\right) - 2 \pi (1+\epsilon \mathcal{T}^n ) {\mathcal{T}'}^{n}\right] 
    + (1-\vartheta ) \left[ \frac{\partial}{\partial Z} \left(\frac{1}{\mathcal{R}^{n+1}} \frac{\partial P^{n+1}}{\partial Z}\right) - 2 \pi (1+\epsilon \mathcal{T}^{n+1} ) {\mathcal{T}'}^{n+1}\right]\\
    - \left[\vartheta  2 \pi (R_{e,0} + \epsilon \beta P^{n} ) + (1-\vartheta) 2 \pi(R_{e,0} + \epsilon \beta P^{n+1} )\right]
    \beta \frac{P^{n+1} - P^{n}}{\Delta T} = 0.
    \label{eq:theta_scheme_pressure_equation}
\end{multline}
Multiplying \eqref{eq:theta_scheme_pressure_equation} by a test function $\psi$ and integrating over the domain $Z\in [0,L]$, we get the weak form:
\begin{multline}
    \int_0^{L} \biggl\{ \vartheta \frac{\partial}{\partial Z} \left(\frac{1}{\mathcal{R}^n} \frac{\partial P^n}{\partial Z}\right) \psi - \vartheta 2 \pi \left(1+\epsilon \mathcal{T}^n \right) {\mathcal{T}'}^{n} \psi \\
    + (1-\vartheta) \frac{\partial}{\partial Z} \left(\frac{1}{\mathcal{R}^{n+1}} \frac{\partial P^{n+1}}{\partial Z}\right) \psi - (1-\vartheta) 2 \pi \left (1+\epsilon \mathcal{T}^{n+1} \right) {\mathcal{T}'}^{n+1} \psi\\
    - \left[\vartheta  2 \pi ( R_{e,0} + \epsilon \beta P^{n}) \psi + (1-\vartheta) 2 \pi ( R_{e,0} + \epsilon \beta P^{n+1} ) \psi \right] 
    \beta \frac{P^{n+1}-P^{n}}{\Delta T} \biggr\} dZ=0.
    \label{eq:weak_pressure_eq}
\end{multline}

\subsubsection{Boundary conditions}

We focus on periodic BCs: $P(0,T)=P(L,T)$ for all $T>0$. Integrating by parts the ``divergence form'' $n$th time stage terms of the weak form \eqref{eq:weak_pressure_eq}, we have
\begin{align}
    \vartheta \int_0^L \frac{\partial}{\partial Z} \left(\frac{1}{\mathcal{R}^n} \frac{\partial P^n}{\partial Z}\right) \psi \,dZ= \vartheta \left[ \frac{1}{\mathcal{R}^n} \frac{\partial P^n}{\partial Z} 
    \psi \right]_0^L 
    - \vartheta \int_0^L \frac{1}{\mathcal{R}^n} \frac{\partial P^n}{\partial Z} 
    \frac{\partial \psi}{\partial Z} \, dZ.
    \label{eq:nth_stage_div_term}
\end{align}
For periodic boundary conditions, $P^n(0) = P^n(L)$, $\psi(0) = \psi(L)$, 
and the flux is periodic, i.e., 
$\left.\frac{1}{\mathcal{R}^n}\frac{\partial P^n}{\partial Z}\right|_{Z=0} = 
\left.\frac{1}{\mathcal{R}^n}\frac{\partial P^n}{\partial Z}\right|_{Z=L}$, 
so the boundary term in \eqref{eq:nth_stage_div_term} vanishes:
\begin{equation}
    \left[ \frac{1}{\mathcal{R}^n} \frac{\partial P^n}{\partial Z} 
    \psi \right]_0^L 
    = \left.\frac{1}{\mathcal{R}^n}\frac{\partial P^n}{\partial Z}
    \right|_{Z=L} \psi(L) 
    - \left.\frac{1}{\mathcal{R}^n}\frac{\partial P^n}{\partial Z}
    \right|_{Z=0} \psi(0) = 0.
\end{equation}
Therefore, the term from \eqref{eq:nth_stage_div_term} reduces to:
\begin{equation}
    -\vartheta \int_0^L \frac{1}{\mathcal{R}^n} \frac{\partial P^n}{\partial Z} \frac{\partial \psi}{\partial Z} \, dZ.
\end{equation}
Similarly, for the $(n+1)$st time stage term:
\begin{equation}
    (1-\vartheta) \int_0^L \frac{\partial}{\partial Z} 
    \left(\frac{1}{\mathcal{R}^{n+1}} \frac{\partial P^{n+1}}{\partial Z}\right) 
    \psi \, dZ 
    = -(1-\vartheta) \int_0^L \frac{1}{\mathcal{R}^{n+1}} \frac{\partial P^{n+1}}{\partial Z} \frac{\partial \psi}{\partial Z} \, dZ.
\end{equation}

Hence, the final nonlinear variational problem to be solved for the pressure is: find $P^{n+1}$ given $P^n$ such that, for every $\psi$,
\begin{multline}
    - \vartheta \int_0^L 2 \pi (1+\epsilon \mathcal{T}^n) {\mathcal{T}'}^{n} \psi \,dZ -\vartheta \int_0^L \frac{1}{\mathcal{R}^n} \frac{\partial P^n}{\partial Z} \frac{\partial \psi}{\partial Z} \, dZ \\
     - (1-\vartheta) \int_0^L 2 \pi (1+\epsilon \mathcal{T}^{n+1}) {\mathcal{T}'}^{n+1} \psi \,dZ -(1-\vartheta) \int_0^L \frac{1}{\mathcal{R}^{n+1}} \frac{\partial P^{n+1}}{\partial Z} \frac{\partial \psi}{\partial Z} \, dZ \\
     - \vartheta \int_0^L  \psi  2 \pi (R_{e,0} + \epsilon\beta P^{n}) \beta \left(\frac{P^{n+1}-P^{n}}{\Delta T}\right) dZ\\ 
      - (1-\vartheta)  \int_0^L \psi 2 \pi ( R_{e,0} + \epsilon\beta P^{n+1} ) \beta \left(\frac{P^{n+1}-P^{n}}{\Delta T}\right) dZ =0.
      \label{eq:final_pressure_weak_form}
\end{multline}

We used the $\vartheta$ scheme with $\vartheta=0.48$ to damp unwanted high-frequency oscillations. The solution of the weak problem \eqref{eq:final_pressure_weak_form} was computed in FEniCS using a first-order Lagrange finite element space. The one-dimensional computational geometry representing the finite axial length of the domain was constructed and meshed using built-in FEniCS functions. The solution is obtained using the \texttt{solve} function with its default settings, which internally calls the \texttt{NonlinearVariationalSolver}.

\subsubsection{Verification and validation}

We performed a verification of the pressure equation's solution by computing the relative errors in the pressure field and cycle-averaged flow rate $\langle Q \rangle$ for the eccentric annular geometry with $ecc=0.2$, $Da=10^{-2}$, and $\beta=100$. Three parameters were verified independently: the one-dimensional mesh size, the number of cycles required to eliminate initial transients, and the number of time steps per cycle. Here, $\langle Q \rangle$ was chosen as the quantity for determining the number of time steps per cycle because, in certain cases where its absolute value was very small, its relative error remained large even after the relative pressure $L^2$ error criterion was 
satisfied. This made $\langle Q \rangle$ the more stringent convergence metric. For the mesh convergence study, we considered three meshes of size $800$, $1600$, and $3200$ elements, and computed the pressure $L^2$ error at $T=\pi/2$ in the last cycle relative to the $3200$-element reference solution, with the total duration and time steps per cycle fixed at $10$ cycles and $12\,800$, respectively. The pressure $L^2$ error decreased from $0.0022\%$ at $800$ elements to $0.0005\%$ at $1600$ elements, confirming convergence, and a mesh size of $1600$ elements was selected for subsequent computations. 

Similarly, for the cycle convergence study, with the mesh size fixed at $1600$ elements and $12\,800$ time steps/cycle, we considered total durations of $5$, $10$, and $20$ cycles, and computed the pressure $L^2$ error (at $T=\pi/2$ in the last cycle) for $5$ and $10$ cycles relative to the $20$-cycle reference solution. The pressure $L^2$ error decreased from $0.0038\%$ after $5$ cycles to $0.0025\%$ after $10$ cycles, confirming that $10$ cycles are sufficient to eliminate initial transients.

Finally, for the time-step convergence study, with the mesh size fixed at $1600$ elements and the total duration fixed at $10$ cycles, we considered time-step resolutions of $6400$, $12\,800$, and $25\,600$ time steps/cycle, and computed the $\langle Q \rangle$ error for $6400$ and $12\,800$ time steps/cycle relative to the $25\,600$ time steps/cycle reference solution. The $\langle Q \rangle$ error decreased from $39.17\%$ at $6400$ time steps/cycle to $13.06\%$ at $12\,800$ time steps/cycle. This relatively large error at $12\,800$ time steps/cycle arises because $\beta=100$ represents the most compliant-wall condition considered, which drives $\langle Q \rangle$ toward a very small absolute value (on the order of $10^{-10}$). This small magnitude amplifies the relative error despite the consistent decreasing trend with increasing temporal resolution. 

Considering the trade-off between accuracy and computational cost, $12\,800$ time steps/cycle was selected to ensure adequate accuracy across all cases. The monotonic decrease in all error metrics for each quantity as resolution increases confirms the convergence of the numerical solution.

To validate the numerical implementation, we compared the finite element solution against the perturbation expansion solution from \eqref{eq: linearized P sol} at $Z=L/4$ during the last cycle. The comparison showed a relative pressure $L^2$ error of $0.52\%$, confirming the close agreement between the numerical and analytical solutions, and validating the finite element implementation.

\end{appen}


\bibliography{references}

\end{document}